# TOWARD LOW EARTH ORBIT (LEO) APPLICATIONS: THE SCIENTIFIC JOURNEY OF THE "SPACE PULSATING HEAT PIPE" EXPERIMENTS

**Marengo M.\*[1,10], Abela M.[2], Araneo L.[3], Ayel V.[4], Bernagozzi M.[1], Bertin Y.[4], Bozzoli F.[5], Cattani L.[5], Cecere A.[6], Filippeschi S.[2], Georgoulas A.[1], Nikolayev V.[7], Mameli M.[2], Mangini D.[8], Mantelli M.[9], Miché N.[1], Pietrasanta L.[1], Romestant C.[4], Savino R.[6], Slobodeniuk M.[4], Toth B.[8], Vincent-Bonnieu S.[8]**

\*Author for correspondence:
Advanced Engineering Centre, University of Brighton, Lewes Road, Brighton, UK,
E-mail: m.marengo@brighton.ac.uk, marco.marengo@unipv.it

[1] Advanced Engineering Centre, University of Brighton, Lewes Road, Moulsecoomb Campus, Brighton, UK
[2] University of Pisa, DESTEC, University of Pisa, Largo Lucio Lazzarino 2, 56122 Pisa, Italy.
[3] Politecnico di Milano, Dept. of Energy, via Lambruschini 4A, Milan, Italy
[4] PPRIME – ENSMA – Université de Poitiers, 1 Av. Clement Ader, 86961 Futuroscope-Chasseneuil, France
[5] University of Parma, Department of Engineering and Architecture, Parco Area delle Scienze 181/A, Parma, Italy
[6] University of Naples, Dipartimento di Ingegneria Industriale, Piazzale Tecchio 80, 80125 Napoli, Italy
[7] CEA/CNRS, Université Paris-Saclay, CEA Saclay, 91191 Gif-sur-Yvette Cedex, France
[8] ESA/ESTEC Keplerlaan 1, Postbus 299, NL-2200AG Noordwijk, The Netherlands
[9] LABTUCAL, Federal University of Santa Catarina, Florianópolis, Brazil
[10] Department of Civil Engineering and Architecture, University of Pavia, via Ferrata 3, 27100, Pavia, Italy

## ABSTRACT

This paper shortly summarises the experimental results obtained since 2011 by a large European academic consortium for the scientific conceptualisation, the definition of the technical requirements, the generation of experimental data, and the validation of a numerical code, for the Pulsating Heat Pipes (PHP) experiment on the International Space Station (ISS). The PHP is a passive, wickless thermal device, whereby a two-phase fluid, forming liquid plugs and vapour slugs, moves with a pulsating or circulating motion inside a meandering tube or channel. The PHP may have a very broad range of geometries (flat, tubular, 3D structured), it can dissipate heat from large areas, and it can be suitable for high power applications with low/medium heat fluxes. PHP functioning is based on the capillary effect, which provides the existence of liquid plugs completely filling the channel cross-section, in a way that any expansion or contraction of the vapour slugs will naturally generate a movement of the fluid along the channel axis. For this, it is important that the channel has a cross-section size below a given threshold, which depends on the liquid surface tension and (for a static fluid) on the gravity acceleration. In space, when only residual accelerations are acting, such a static size threshold is virtually infinite, while a finite dynamic threshold exists even in the absence of gravity. The concept of a "Space PHP" was originally developed in 2014 by the team, and from then 17 Parabolic Flight Campaigns (PFC) and 3 Sounding Rocket (SR) experiments have been carried out to generate the data for the preparation of an experiment targeting a Low Earth Orbit (LEO) mission. Both a tubular and a flat plate PHP have been successfully tested in reduced gravity and on ground, by using different combinations of fluids and building materials. The need for having an experiment on a LEO environment is mainly because, during a PFC, only 22sec of reduced gravity are possible, which is a period below the characteristic time for reaching a steady state condition for almost all of the tested devices.

Instead, a steady state was reached using the SR campaigns: in this case however, only one experimental condition was achievable, and long-duration data of the PHP performance still remains beyond reach. Several measurement methodologies have been used to characterise the Space PHP, like infrared analysis, high-speed camera visualisation techniques, with data processed with different techniques, from wavelets to inverse heat transfer problem solution. The results clearly showed that PHPs are very interesting for space applications due to their simplicity of construction, the capacity to transfer heat up to several hundred watts, a high power/weight ratio, their geometrical adaptability, and, in particular, the Space PHP will be a breakthrough technology for space thermal management.

## 1. INTRODUCTION

Two-phase passive heat transfer devices have an important role in space thermal control systems, still gaining more and more relevancy because of their lightweight, high performances and reliability. When heat dissipation becomes very high, heat pipes are preferred to honeycomb structures in radiators panels and they have been widely used to reduce temperature gradients. The Capillary Pumped Loops (CPL) and Loop Heat Pipes (LHP) have been successfully used as thermal control systems in a large number of space missions because of their ability to cover tortuous and longer paths. Pulsating heat pipes (PHP) have been widely studied in the last 15 years, even if their concept was presented already 30 years ago by Akachi [1]. In particular, since more than 10 years, the authors have investigated the use of PHPs for space application, i.e. in a microgravity environment. In particular, they developed the new concept of a "Space PHP" which profits of the absence or a lower gravity acceleration to achieve higher performances in terms of higher heat fluxes, higher thermal power loads and lower thermal resistances. A PHP is made of a wickless capillary channel or tube bent in several turns between a heated and a cold zone. The tube is partially filled with a working fluid at the liquid/vapour saturation state, which is internally distributed by means of surface tension effects in the form of liquid slugs and vapour plugs. At the evaporator side, under a given heat flux, liquid film evaporation or flow boiling occur. At the condenser side, vapour condense toward the liquid film on the channels' walls. Flow patterns ranging from slug flow to annular flow are initiated in adjacent tubes by local pressure instabilities, and they effect the global heat flux transferred from the heated to the cooled region. When the flow patterns are considered in conjunction with related heat transfers, many parameters have a direct influence on PHP operation; while the most important is obviously the channel internal diameter that allows liquid/vapour phase division into liquid slugs and vapour plugs separated by menisci, other parameters are the number of turns, PHP length, filling ratio, the physical properties of the working fluid or, indirectly, the cold source temperature, heat power applied or heat flux, the inclination with respect to gravity and the closure, or not, of the overall loop. We address the reader to our reviews of the field for more information about the PHP principles and operations [2,3]. A PHP is working on the principle that a liquid slug fills completely the channel when its hydraulic diameter is below a critical capillary diameter. If we are not considering dynamic effects, a tube can be considered "capillary" if the surface tension forces overcome the gravity forces, maintaining the internal liquid in a confined mode. This concept can be expressed using the dimensionless number Bond $Bo = \rho g D^2 / \sigma$: When the Bond number is below one, the surface tension dominates, and therefore we are in the conditions written above, i.e. the liquid is not stratified and fills the channel. As we will see afterwards [4], the critical threshold is not exactly equal to the unity, and varies with the liquid slug acceleration, wall wettabilities and other parameters. It is evident that, when the gravity acceleration is very small, the $Bo$ number tends to be very small, i.e. the critical diameter tends to increase to very high values. In space, even big channels are capillary, in the sense described above. During the 58th ESA PFC in 2013, Mameli et al. [5] carried out a first experimental investigation on a Capillary Loop PHP tested both on the ground and under micro/hyper gravity conditions. This PHP consisted in a copper tube with an internal diameter of 1.1 mm (just under the critical diameter for FC-72, calculated with the static criterion based on the Bond Number at 1g, $D_{crit}$ =1.68 mm at 20ºC), bent into a planar serpentine, for a total of 16 U-turns in the evaporator zone. Tests during the PFC showed that the vertical operation was affected by hyper- or microgravity phases, the first of them slightly assisting the flow motion, and the second one leading to a sudden increase of temperature in the evaporator zone. Such behaviour was quite similar to the thermal dynamic response of vertical-to-horizontal tilt manoeuvre on ground. The horizontal operation does not seem to be affected by variations of the gravity field.

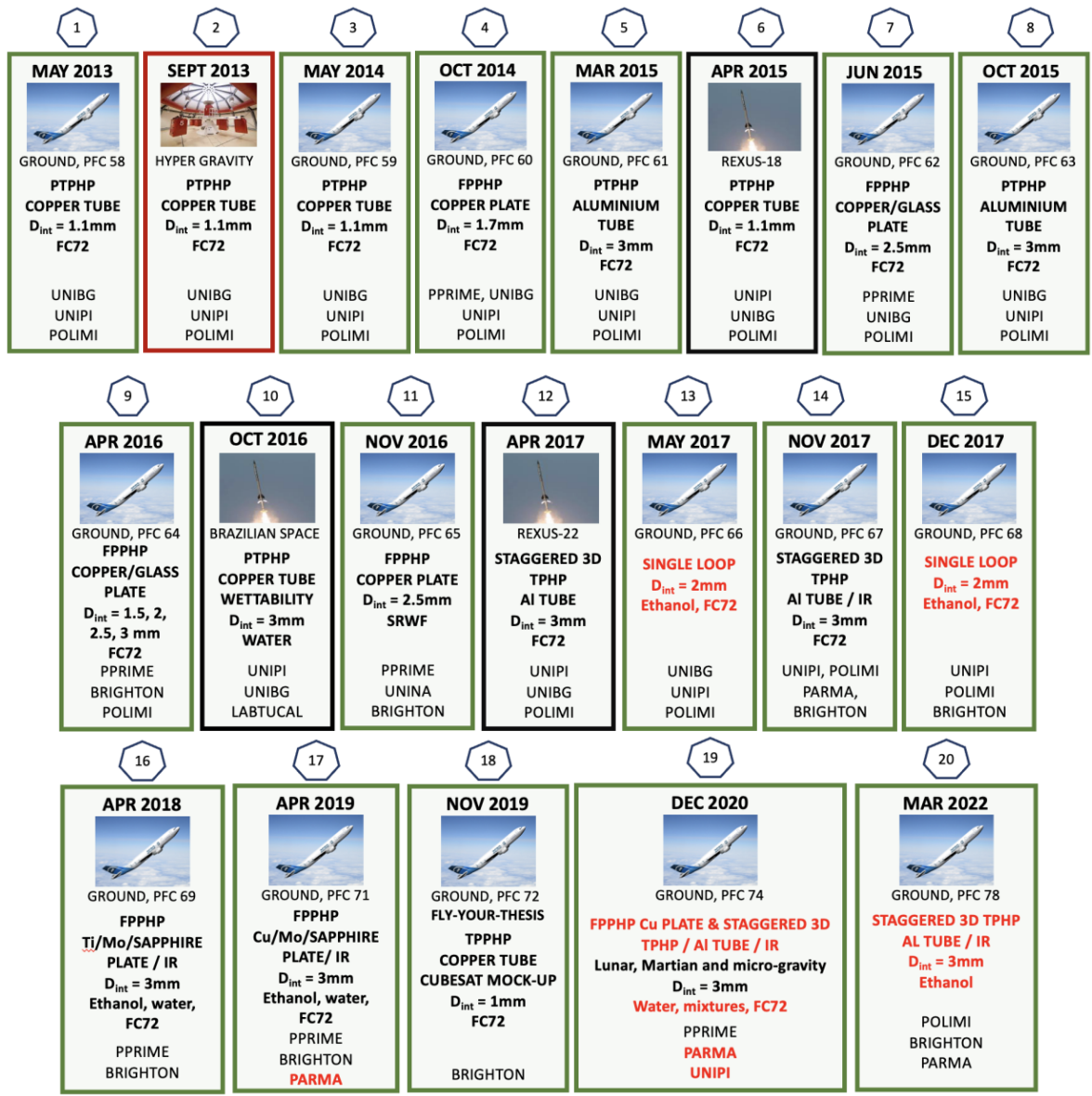

**Fig. 1** A graphical representation of the most important research activities carried out by the research team in micro-, terrestrial and hyper-gravity from 2013. In this paper we will describe results obtained only in 7 campaigns, leaving the reader to consider the appropriate literature for more details.

In 2003, also Gu et al. [6] experimented two flat plate aluminium PHPs (250 x 60 x 2.2 mm$^3$) in PFC each with three different configurations. Each PHP was filled with R114 (FR 50%). A continuous channel, engraved on the bottom plate, had a cross section of 1x1 mm$^2$ and was laid out with 48 turns at both ends of the closed loop. The evaporator consisted of a 55x16 mm$^2$ Kapton heater, and for the condenser a fan of 80x80 mm$^2$ was used: in one configuration the evaporator is at the centre of the PHP, while for the other two, the heater is at the bottom (BHM) and at the top (THM), respectively. The authors observed that both PHPs operated more effectively in all the configurations under reduced gravity than under normal or hypergravity phases. Based on a dimensional analysis, these authors were the first to consider the use of larger channel diameters for space applications.

At the moment, two experiments were able to carry out a long duration test of a PHP in space, one from Japan Aerospace Exploration Agency (JAXA) in cooperation with Tohoku University in Japan, and the other one from the Air Force Research Laboratory partnered with Thermavant Technologies and Maxar Technologies (USA). Both experiments are with a standard capillary size. The on-orbit experiment of JAXA considered a stainless steel 5 turns 2D TPHP with $D_{int}$=0.8 mm, filled at 45% with HFC-134a, and has been conducted since 2012 as one of the Small Demonstration Satellite-4 (SDS-4) missions [18]. The particularity of this experiment has been the use of a check-valve to promote the flow circulation. The US experiment intended to demonstrate the functionality and the limits of operation of a PHP [19]. The experiment was installed on the 5[th] flight of the Air Force's X-37B spaceplane and lasted 780 days until Oct 2019. The Advanced Structurally Embedded Thermal Spreader (ASETS-II) flight experiment consisted in three 2D TPHP ($D_{int}$ = 1mm, 34 channels) with different varying configurations (centre heating with large and small heaters and with single- and double-ended cooling), using butane and R-134a. During six-week continuous operation test, the PHPs performed as expected with negligible differences between horizontal and microgravity flight data.

As written above, both Gu et al. [6] and our research group [5, 8] illustrated the possibility to build a PHP for space application with an internal diameter bigger than the static critical diameter on ground. Under reduced gravity, body forces are negligible and the threshold diameter to obtain a slug-plug configuration increases. Since the mass of the thermal fluid per unit length is proportional to the square of the $D_{int}$, increasing the inner diameter is also beneficial in terms of total heat exchanged. Theoretically, for negligible gravity accelerations, the capillary diameter tends to infinite, however, the limit to the increase of the inner diameter is also given by inertial and viscous effects. The whole work has been always related to the concept of confined flow, i.e. the condition of having the liquid slugs filling completely the channels, in order to trap the vapour, which is acting as a thermal pump during evaporation and condensation. Therefore, the classical use of the Bond Number cannot work for microgravity conditions, since the Bond number is always zero for any channel size. The inertial and viscous terms are becoming very important with respect to the capillary phenomena, and authors in literature used dimensionless number such as Weber and Reynolds.

In two papers of the research group [4, 16], the problem has been deeply analysed through the data collected on ground and in microgravity, where the gravity acceleration should not be considered inside the correlations. It resulted that the use of the actual bubble acceleration and the bubble length, leading to the constructions of modified dimensionless numbers, is more efficient in distinguishing the various PHP regimes. Despite the fact that it is not possible to use such criteria for the design of the PHP, since bubble length and acceleration are not known a priori, those limits can be implemented in lumped numerical codes for the simulation of Pulsating Heat Pipes [19] or in VOF-CFD models, to eliminate the empirical assumptions on the existence of a slug-plug flow pattern.

## 2. TOWARD THE SPACE PULSATING HEAT PIPE

This paper summarises some of the results obtained by the authors in the last 10 years on the development of the concept and characterisation of the so-called Space Pulsating Heat Pipe. In Figure 1, it is possible to visualise all the microgravity experiments carried out until now by the research group on this technology.

### 2.1 The Planar Tubular Pulsating Heat Pipe

A first hybrid Thermosyphon/Pulsating Heat Pipe with a diameter bigger than the terrestrial capillary limit was tested both on ground and in hyper/micro gravity conditions during the 61st ESA PFC in 2014 [7]. During each of the 3 flight days, 31 parabolic trajectories are performed: the first one, called parabola zero, is followed by six sequences of five consecutive parabolae. In between each sequence, there are five minutes of steady flight at normal gravity level, used here to vary the heat power level and let the system reach a steady state when possible. Each parabola is then composed by a first hyper-gravity period (20s at 1.8g), the micro-gravity period (20s at 0g) and a second hyper-gravity period equal to the first. This first device consisted in an aluminum tube ($D_{int}$ = 3 mm) bent into a planar serpentine with five curves at the evaporator zone, while a transparent section closes the loop, allowing fluid flow visualizations in the condenser zone (see Fig. 2). Five heaters, mounted alternatively in the branches just above the curves at the evaporator zone, provide a possible asymmetrical heating thus promoting the fluid flow circulation in a preferential direction [8].

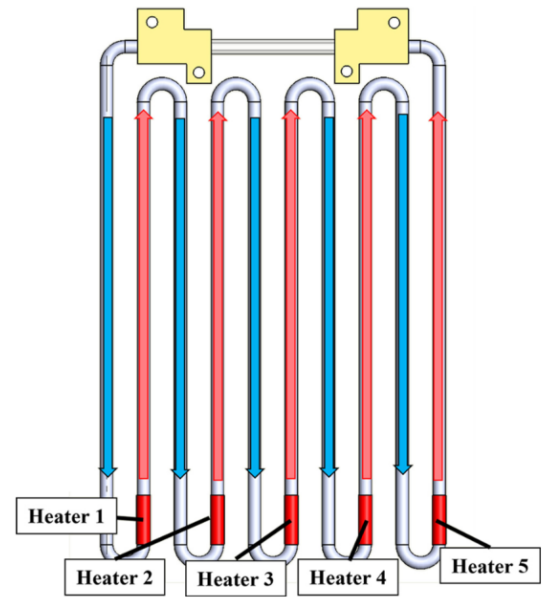

**Fig. 2** The first Planar Tubular Pulsating Heat Pipe with 5 turns analysed in 2014, with 5 independent heaters.

On ground, the device is first vacuumed by means of an ultra-high vacuum system down to 0.3 mPa, and then it is partially filled up with the working fluid (FC-72) with a volumetric ratio of 50% ± 3% and sealed by means of tin soldering. The fluid itself is previously degassed within a secondary tank, by continuous boiling and vacuuming cycles. Presence of NCG in the system was detectable only indirectly through verification of the pressure value, which should correspond to the saturation pressure at the given temperature. This kind of procedure is common to all the tests, and it will be not repeated anymore in the paper. The device has been tested at different positions (vertical and horizontal) and at different heat power input levels (from 10 W to 160 W). Ground tests showed that this device works as a thermosyphon when gravity assisted: in vertical position the device can reach an equivalent thermal resistance of 0.1 K/W with heat fluxes up to 17 W/cm². In horizontal position the fluid motion is absent, thus the device works as a pure thermal conductive medium. In Fig. 3a and 3b, it is evident that the hybrid PHP is working better in vertical BHM position, achieving 180W of heat power, still with less than 70°C for the evaporator temperature. The PHP is still working in the horizontal position, and it is possible to verify that the thermal performance at microgravity resembles the ones at microgravity (Fig. 3b & c), and that in microgravity the liquid slugs are filling completely the tube as expected (Fig. 3d). Pseudo-steady state conditions can be reached in approximately 3 min, due to the low thermal inertia of the heating system and the device: In such a sense, during the parabolas the system never reached its steady state. In conclusion, these parabolic flight tests reveal that the device operates in a completely different way when the microgravity is reached: the images recorded in the condenser zone, together with the pressure signal shows a transition from the Thermosyphon mode to the PHP working mode.

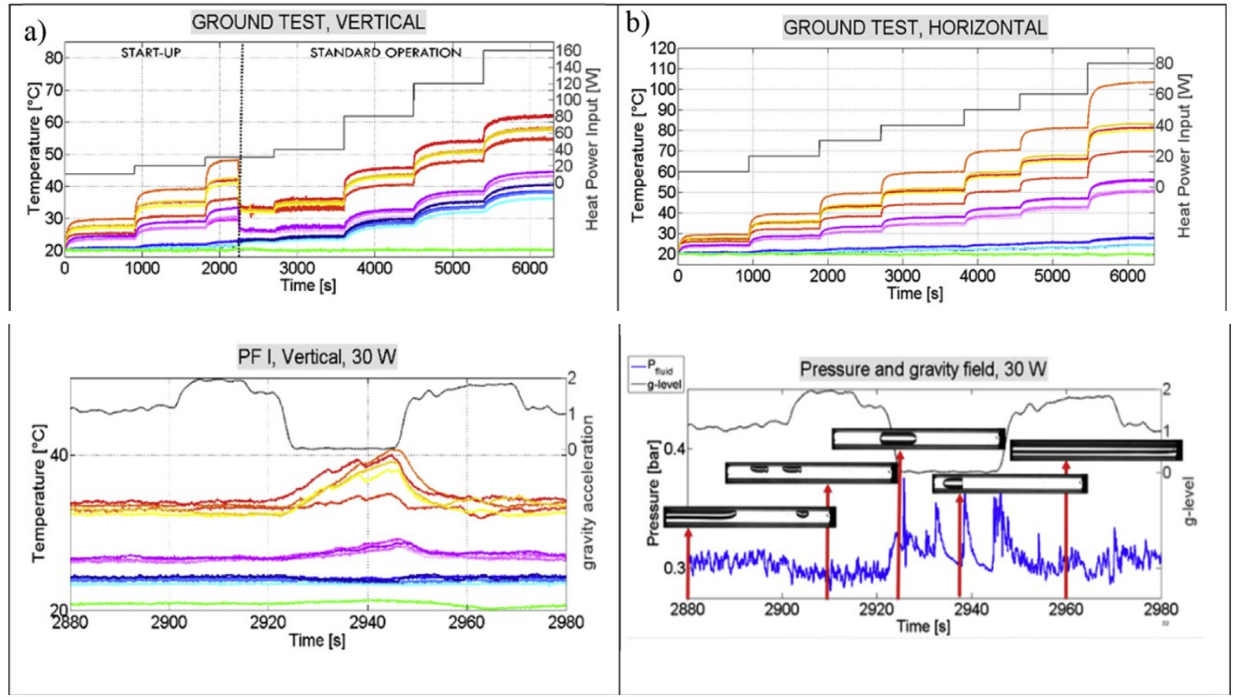

**Fig. 3** A standard visualization of the temperature measurements and the fluid flow for the 2D Planar TPHP. In (a) and (b) the ground tests, in (c) and (d) one exemplary parabola. The lines with the different colours in (a) and (c) represent the values from the 14 thermocouples at the evaporator (reddish) and condenser (bluish and greenish). The hypergravity and microgravity zones are recognizable by looking at the gravity acceleration measurements (line in black in (c) and (d). In (d) the values of the fluid pressure are synchronised with the accelerometer and the visualizations of the high speed grey scale camera.

In some cases, the hyper-gravity period is able to eliminate partial dry-outs restoring the correct operation until the occurrence of the next microgravity period. A following experiment on the same kind of hybrid plan tubular PHP has been carried out in 2015 (63rd ESA PFC) considering the idea of different heating zones [8]. In most of the space applications, the heating can be localised in given positions linked to presence of CPU, IGBT and other electronic components. Moreover, it is possible that the switch on of heat fluxes with different patterns may facilitate the working operation of the PHP in circulation mode, increasing the thermal performances. For this experiment, the bubble velocity was estimated with an open source Particle Image Velocimetry (PIV) software partially modified in order to detect the liquid/vapor interface motion along the transparent section. The bubble velocity is calculated utilizing the OpenPIV software. With respect to the hybrid planar PHP with uniform heating, when the device is heated in a non-uniform way, a circulation in a preferential direction occurs, with an improvement of the overall thermal performance, especially for the lower heating power tested. In microgravity conditions, the sudden absence of the buoyancy force permits to activate a typical PHP oscillating slug/plug flow. Non-uniform heating configurations have a beneficial impact also in microgravity. Increasing the heating power level at the most lateral branches and decreasing it at the centre, the stop-over periods observed in microgravity for the uniform heating, are limited. In this way, the possibility to stabilize the two-phase flow motion in weightlessness conditions was proved through different heating configurations at the evaporator.

## 2.2 The Staggered 3D Tubular Pulsating Heat Pipe

A completely new device has been proposed in Nov 2017 for the 67th ESA PFC. It is made of an annealed aluminum (6060 alloy) tube with an inner and outer diameter of 3 and 5 mm, respectively. The closed loop is folded in a staggered 3D configuration with 14 turns (8 mm bending radius) in the evaporator zone, as shown in Fig. 4. The geometry (overall size of the PHP: $220 \times 80 \times 25$ mm) was selected to fit this device in a Heat Transfer Host designed for the experiment on the International Space Station. Two brass connections allow to connect a transparent sapphire tube with the aluminum tube, and to host two K-type micro-thermocouples for the fluid temperature measurement (red and blue "F" bullet in Fig. 4b), as well as another miniature pressure transducer close to the evaporator section (green "P" bullet in Fig. 4b). Sapphire is chosen for its optical properties, being almost transparent also in the MWIR spectrum The heating power is provided by a programmable power supply from a minimum of 18 W to 182 W (1.10 to 11.43 W/cm$^2$).

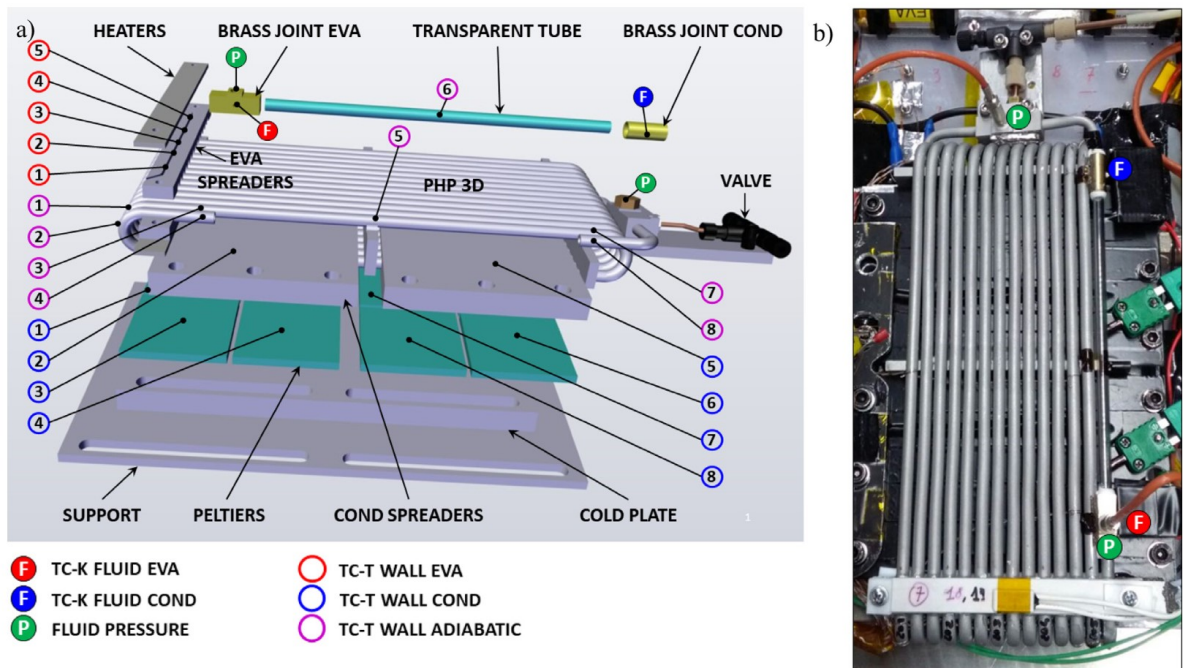

**Fig. 4** The 3D staggered TPHP: thermocouples and pressure transducer locations: (a) the CAD exploded view, (b) the actual front view.

The condenser zone is embedded between two aluminium heat spreaders cooled down by means of an array of 8 Peltier cells (82.2 W each). A single-phase auxiliary cooling loop has been used to dissipate the waste heat of the Peltier cells, coupling the hot side of each element with a cold plate. Five T-type thermocouples are located between the evaporator spreader and the heater; six are located between the Peltier cold side and the condenser aluminium heat spreader; two on the condenser heat spreader just behind the sapphire tube; seven are located on the tube external wall (Fig. 4a). The PHP is filled with FC-72 (FR 50%). In comparison with a previous experiment with a smaller number of heated sections [7, 8], the absence of stop-over periods is beneficial in terms of heat transfer rate. Indeed, for the same wall-to-fluid heat input flux (the effective heating area passed from $A_{eff} = 9.4$cm$^2$ to 15.8cm$^2$) this 3D PHP is characterized by smaller transient temperature differences for all the heating powers. Start-up tests, where the heat load is provided just after the occurrence of microgravity, show that the 20s microgravity are enough for the device activation: the PHP operation is not primed by the flow inertial effect, still present when the device is activated before the micro-gravity period. The simultaneous measurement of the internal fluid pressure and temperature has been successfully implemented showing the existence of subcooled and superheated local thermodynamic states.

## 2.3 The Flat Plate Pulsating Heat Pipe

Another kind of PHP tested by the research group was the so-called Closed Loop Flat Plate Pulsating Heat Pipe (FPPHP) (Fig. 5) tested on ground and on board of an aircraft during the 60th ESA PFC in 2014 [10]. The purpose was to test an innovative FPPHP, which behaves differently from tubular PHPs due to two main characteristics: (i) the square/rectangular channels with sharp angles and (ii) the transverse heat spreading throughout the continuity of the plate bulk. The FPPHP consists of two brazed copper plates, into one of which a continuous rectangular channel (1.6 x1.7 mm$^2$) with 12 turns in the evaporator, is machined.

The channel is filled with FC-72 as working fluid (FR 50%). Tests have been conducted with the FPPHP positioned both horizontally and vertically in BHM. The FPPHP presented an innovative design, involving the milling of grooves between the channels to avoid the heat spreading across the channel (Fig. 5b). The PHP and the heat sink are clamped against one another with through-bolts traversing nine screw holes. Experimental results on the ground show that the thermal device can transfer more than 180 W in both inclinations, and that the horizontal operation is characterized by repeated stop-and-start phases and lower thermal performance. The FPPHP can operate under microgravity conditions and with a transient gravity force, with global thermal resistance reaching 50% and 25% of that of the empty plate, in horizontal and vertical orientation respectively. A minimum thermal resistance of 0.12K/W was recorded. Compared to the tubular PHP tested by Mameli et al. [7, 8], this FPPHP has proved to be quicker to respond to gravity variations in horizontal inclination, mainly on account of the higher hydraulic diameter of its channels. The FPPHP tested in horizontal inclination is not influenced by changes of gravity levels, proving that such a system can be an efficient candidate for thermal control in different space applications, once a greater channel size is combined with a sufficient number of turns. Another FPPHP, again filled with FC72, was tested during the 62$^{th}$ ESA PFC under vertical orientation in 2015 [11]. The FPPHP is made of a thin copper plate, in which a curved channel disposed with 11 U-turns is milled and closed on the top face by a transparent borosilicate plate. The particular characteristics of this FPHP is that the equivalent hydraulic diameter of the square channel (2.5×2.5 mm$^2$) is above the working fluid capillary diameter on ground. On ground, the heat transfer mode is represented either by pure pool boiling or by an annular flow pattern inside the channels mostly filled by the refrigerant vapor. Instead, during microgravity, the fluid regime naturally turns into a slug-plug flow pattern. During the transition from 1.8 $g$ to 0 $g$ a rapid dry-out may occur in some of the channels, followed by a similarly fast reaction of liquid plugs moving towards the evaporator from the condenser zone. Such stop-and-start motion events continue during the whole microgravity period, leading to strong temperature oscillations, but also to a still acceptable thermal performance of the device. The same FPPHP was tested during the 65$^{th}$ ESA PFC under vertical orientation in 2016 [12], but filled with pure water and a self-rewetting mixture (water/butanol), whereby for the last one, the surface tension increases with temperature in a small range around 60ºC, which may lead to a rewetting of the heated zones. Indeed, for low power input, the FPPHP filled with pure water was not able to work under low-g conditions, because dry-out condition appears in some channel, while the FPPHP filled with SRWF was able to operate. Furthermore, its thermal resistance was lower [12]. Four similar prototypes of FPPHP were also tested during 64th ESA PFC in 2016 [13], with square channel dimensions ranging from 1.5 mm to 3 mm. It was observed that thermal performances increase with increasing diameter, both in vertical BHM and microgravity transient phases. The larger the channel diameter, the higher the number and cumulated time of reactivation or active phases, up to a limit that seemed to be situated between 2.5 mm and 3 mm of diameter. This suggests that there exists an optimal hydraulic diameter in such a configuration beyond which performance will not necessarily increase, due to higher inertial forces and/or the lower capillary forces necessary to maintain the menisci in slug/plug distribution.

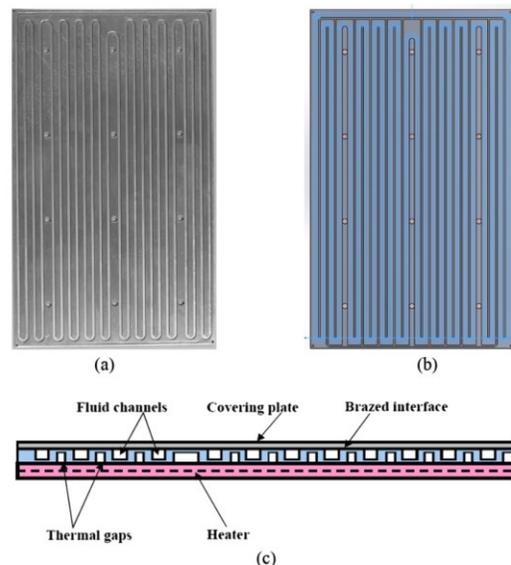

**Fig. 5** The first FPPHP used in a PFC in 2014: (a) the engraved channel on the copper plate, (b) the blue silicon glue used to separate the channels and to seal the upped plate, (c) the section view with the thermal gaps to decrease the lateral heat flow connection.

## 3. DATA ANALYSIS

### 3.1 Infrared Analysis

The infrared analysis is essential for the characterisation of a PHP, not only because it is allowing to have a measure of the metal surface temperature or, in case of the sapphire insert, of the fluid inside the system, but because it is opening the route for the estimation of the heat transfer coefficients, which are very important for the tuning of numerical simulations. There are rare research works on the evaluation of internal flow heat transfer in presence of an oscillation. The research group has first tested the use of the inverse heat conduction problem solution on a Single Loop Pulsating Heat Pipe (Fig. 6) partially filled with ethanol (FR 60%) and designed with two sapphire tubes as a connection between the evaporator and the condenser [14]. The local heat flux was estimated for the flow patterns usually observed during the PHP operation: annular flow and oscillating slug/plug flow. In the region between the evaporator and the condenser, the local wall-to fluid heat flux can reach values of up to 4000 W/m$^2$ in the case of annular flow and 2000W/m$^2$ in the case of an oscillating slug/plug flow. The same technique has been used for the Staggered 3D Tubular Pulsating Heat Pipe [9, 15] during the PFC67 in 2017.Two regimes were identified: intermittent flow (sporadic fluid motion, occurring in some channels) and full activation (stable fluid motion within the whole adiabatic section). During the intermittent flow regime, only some channels present fluid oscillations mainly in the frequency range 0.5–0.6 Hz; on the contrary, for high power inputs to the evaporator, the dominant frequency is generally similar in the whole adiabatic section and it is around 1.0 Hz, even though the oscillatory behaviour may still vary from channel to channel. This dominant frequency values are in good agreement with the ones estimated with pressure sensors by Perna et al. [17]. The 80th percentile of the wall-to-fluid heat flux amplitude is almost constant (about 500 W/m$^2$) at low power inputs. From 100 W up to the maximum power input, the heat flux amplitude increases almost linearly with the power input, reaching about 1500 W/m$^2$ at 202 W.

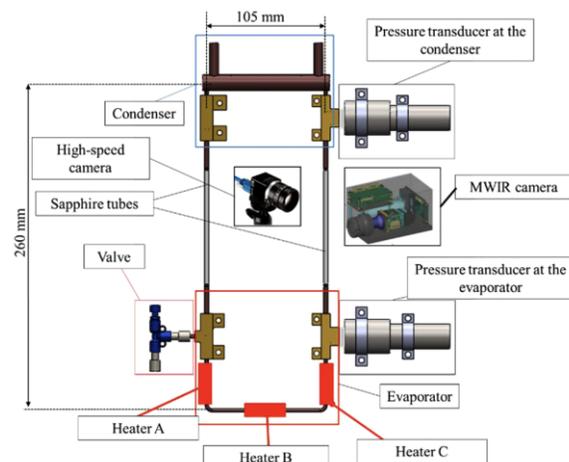

**Fig. 6** The Capillary Single Loop TPHP which served for the first experiments about the evaluation of the heat transfer convection coefficient using a high speed MWIR camera and the resolution of the Inverse Heat Transfer Problem.

## 3.2 Wavelet analysis

The research group has also used the wavelet transform to characterize the fluid pressure signal in the frequency domain varying the heat power input at the evaporator and in the condenser zone of a full-scale PHP tested in microgravity conditions in the 67th ESA PFC [17]. During the slug-plug flow regime, the dominant frequencies falls in the range 0.6–0.9 Hz, showing an increasing trend with the heat load input. The Cross-Correlation reveals that the two signals at the evaporator and at the condenser are very similar.

## 4. CONCLUSIONS

Finally, the long journey of this collective work has proven that the concept of a Space PHP is viable for space applications, both in the Tubular and in the Flat Plate configuration, with a clear indication that more than 10 W/cm$^2$, less than 0.1K/W thermal resistance, more than 450 W/kg, can be achieved with a total power that reached 200W in some configurations. The systems have low start-up powers (from 30W), fast (below 20s reaction time, depending on thermal mass), and they are very reliable in terms of duration (until now a TPHP worked for longer than 1 year in laboratories of Pisa and Poitiers). It is also possible to design a specific Space PHP using a validated lumped parameter code (CASCO by CEA/CNRS). The team is now looking forward to being able to carry out a long-duration experiment in space in the next few years, as it has already been done for the capillary PHP by JAXA and AFRL teams. This work is also extremely important for the future improvement of the status-of-art PHP simulation code CASCO (French abbreviation for "*Code avancé de simulation de caloduc oscillant*"). In a very recent paper [20], CASCO code prediction ability was validated against the experimental data collected during the micro-gravity tests of 3D Staggered TPHP (67th ESA PFC). The advanced 1D Lumped parameter network code was able to predict the temperature temporal evolution of evaporator and tube walls with a maximum deviation of 7%, the pressure trends were reproduced, like the start-up time.

## ACKNOWLEDGMENTS

This research has been a coral activity involving 7 academic European universities and the European Space Agency. Including M.Sc. and PhD students, 87 researchers have been involved in the construction of 19 PHP devices, tested in 21 ESA PFCs and 3 SR missions (https://eea.spaceflight.esa.int/portal/). It is worth also to acknowledge that, within this research framework, three student teams in Pisa and Brighton have been awarded for the Fly-Your-Thesis ESA and Student REXUS SR program. The work has been financially supported by UK EPSRC through the grant EP/ P013112/1 (HYHP), as well as the ESA MAP Projects INWIP and TOPDESS, Emilia-Romagna Region (POR-FESR 2014-2020 (NANOFANCOIL), the financial support from CNES in the framework of the GDR/MFA. Further acknowledgements should be given to Prof. P. Di Marco (UNIPI, Italy), Prof. K. Vieira de Paiva (UFSCA, Brazil), Prof. Sara Rainieri (UNIPR, Italy), Giuseppe Correale (ESA), Prof. J. Bonjour (INSA, Lyon, France), Prof. S. Khandekar (IIT, Kanpur, India) and Prof. S. Van Vaerenbergh (ULB, Belgium).

## NOMENCLATURE

| | |
|---|---|
| BHM : Bottom Heated Mode | FR: Filling Ratio |
| THM: Top Heated Mode | VOF: Volume-of-Fluid |
| PFC: Parabolic Flight Campaign | NCG: Non-condensable gas |
| SR: Sounding Rocket | Bo: Bond Number |
| LEO: Low Earth Orbit | D: diameter |
| $D_{int}$: Internal Diameter | $\rho_l$: liquid density |
| ISS: International Space Station | g: gravity acceleration |
| ESA: European Space Agency | $\sigma$: surface tension |